# A Network-Based Biomarker of Morphological Disruption Associated with Breast Cancer Malignancy

Reza Bozorgpour

College of Engineering and Applied Science, Department of Biomedical Engineering, University of Wisconsin-Milwaukee, Milwaukee, WI, USA

*Corresponding author
E-mail: bozorgp2@uwm.edu

## Abstract

Breast cancer diagnosis commonly considers individual nuclear morphological characteristics, whereas their joint organization is less frequently evaluated. We developed the Morphological Network Disruption Index (MNDI), a patient-level measure of morphological abnormality relative to a benign reference state. Using the Wisconsin Diagnostic Breast Cancer dataset, ten nuclear characteristics were represented as network nodes, with MNDI summarizing disruption across 45 pairwise feature configurations. Performance was evaluated using repeated stratified five-fold cross-validation. Malignant lesions exhibited substantially higher MNDI than benign lesions (mean: 3.454 vs. 1.165; $p = 2.94 \times 10^{-69}$). MNDI achieved an AUC of 0.941 (95% CI: 0.918–0.961), with 85.9% sensitivity and 91.6% specificity. A classifier using network-derived descriptors achieved an AUC of 0.928 ± 0.024. Independent evaluation in the BreaKHis histopathology cohort showed limited discrimination (AUC = 0.538, 95% CI: 0.401–0.668), indicating dependence on the underlying morphological representation. MNDI provides an interpretable framework for quantifying patient-specific morphological disruption, while individual implementations require validation within compatible feature spaces.

## Introduction

Breast cancer is a heterogeneous disease characterized by substantial variation in morphology, molecular features, biological behavior, clinical presentation, and therapeutic response [1,2]. Despite major advances in molecular classification and precision oncology, histopathological assessment remains fundamental to breast cancer diagnosis and classification [3,4]. Evaluation of hematoxylin and eosin (H&E)-stained tissue provides essential information on tumor type and histological grade, while additional pathological characteristics, including lymphovascular invasion, lymph node involvement, and margin status, contribute to prognostic assessment and clinical management. Morphological classification therefore remains an important component of breast cancer characterization alongside immunohistochemical and molecular profiling [5–10].

Histomorphological assessment of breast tumors relies on cellular and architectural characteristics evaluated both individually and in combination. Tumor differentiation can be characterized morphologically through histological type and grade [11,12]. In particular, the Nottingham grading system incorporates three biologically relevant morphological features: tubule or gland formation, nuclear pleomorphism, and mitotic count [13]. The inclusion of nuclear pleomorphism as a central component of histological grading underscores the importance of nuclear morphology in breast cancer characterization and provides a rationale for its quantitative investigation [14].

Quantitative analysis of nuclear morphology has increasingly been explored as a complementary approach for breast tumor characterization. Benign and malignant breast lesions exhibit measurable differences in cellular and nuclear morphology, motivating the investigation of nuclear size and shape characteristics as potential biomarkers of malignancy [15–21]. Machine-learning approaches have further demonstrated that quantitative morphological features can provide substantial discriminatory information for breast cancer classification [22–39]. In a recent analysis of the WDBC dataset, ten nuclear morphological characteristics including measures of nuclear size, shape, texture, and boundary irregularity were systematically evaluated using statistical and machine-learning approaches, demonstrating their utility for distinguishing benign from malignant breast tumors [40–46].

Despite the demonstrated predictive value of nuclear morphological features, most existing computational approaches evaluate these characteristics individually or combine them as inputs to statistical and machine-learning models [47,48]. Such approaches are effective for identifying informative features and predicting malignancy but provide limited insight into the extent to which a patient's joint morphological configuration deviates from a benign reference state. Because morphological characteristics are interrelated, considering their pairwise configurations may provide a complementary representation of tumor-associated morphological abnormality. A patient-specific measure of such deviation could therefore provide an interpretable summary of morphological disruption beyond the assessment of individual features alone.

To address this gap, we developed the Morphological Network Disruption Index (MNDI), a patient-specific framework for quantifying morphological abnormality relative to a benign reference state. Rather than considering nuclear characteristics solely as independent predictors, MNDI represents morphology through pairwise feature configurations and quantifies the extent to which each patient deviates from corresponding benign joint feature distributions. The framework was evaluated using the WDBC dataset to determine whether MNDI could distinguish benign from malignant lesions, assess its stability across repeated held-out evaluations, and identify morphological feature pairs most strongly associated with malignancy-related disruption. In addition, the generalizability of the MNDI framework across morphological representations was examined in the independent BreaKHis histopathology cohort using quantitative nuclear features extracted from segmented H&E images.

# Methods and Materials

## Dataset and Morphological Feature Selection

This study utilized the WDBC dataset, which contains quantitative morphological measurements derived from digitized images of fine-needle aspirate (FNA) samples obtained from breast masses. The dataset includes 569 samples, comprising 357 benign and 212 malignant cases. Each sample is characterized by 30 quantitative variables derived from ten fundamental nuclear characteristics: radius, texture, perimeter, area, smoothness, compactness, concavity, concave points, symmetry, and fractal dimension. For each characteristic, the dataset provides the mean, standard error, and worst value.

The objective of this study was to determine whether breast cancer malignancy is associated not only with alterations in individual morphological characteristics but also with disruption of the relationships among these characteristics. To reduce redundancy while retaining the primary morphological information, only the mean values of the ten nuclear characteristics were used for network construction. For patient $p$, the morphological feature vector was defined as

$$X^{(p)} = [x_1^{(p)}, x_2^{(p)}, \dots, x_{10}^{(p)}]$$

where $x_i^{(p)}$ represents the value of morphological characteristic $i$ for patient $p$.

Because the selected morphological variables differ substantially in their numerical scales and units, feature-wise standardization was performed before network construction [49]. To prevent information leakage, standardization parameters were estimated exclusively from the training data within each cross-validation iteration [50]. For morphological characteristic $i$, the standardized value for patient p was calculated as

$$z_i^{(p)} = \frac{x_i^{(p)} - \mu_i^{(train)}}{\sigma_i^{(train)} + \epsilon}$$

where $\mu_i^{(train)}$ and $\sigma_i^{(train)}$ denote the mean and standard deviation, respectively, of characteristic $i$ in the training data, and $\epsilon$ is a small positive constant introduced to ensure numerical stability. The same training-derived parameters were subsequently applied to the corresponding features in the test data. This procedure placed the morphological characteristics on comparable scales while maintaining complete separation between model development and evaluation data.

## Construction of Patient-Specific Morphological Networks

Each tumor was represented as a patient-specific weighted network in which the ten standardized morphological characteristics constituted the network nodes. Rather than defining an edge simply as the numerical distance between two characteristics, the proposed approach quantified how unusual each pair of characteristics was relative to its joint distribution among benign tumors. This formulation was designed to capture disruption of the relationships normally observed between morphological characteristics.

For every pair of morphological characteristics $i$ and $j$, a two-dimensional benign reference distribution was estimated using only benign samples contained in the corresponding training folds. For a benign training sample b, the pairwise feature vector was represented as

$$q_{ij}(b) = [z_i(b), z_j(b)]$$

The center of the benign distribution for feature pair $i, j$ was estimated from the benign training samples as

$$\mu_{ij}(B) = mean\left[q_{ij}(b)\right], for\ b \in B(train)$$

where $B(train)$ represents the set of benign samples in the training data.

The covariance structure of each benign feature-pair distribution was also estimated from the benign training samples. Because stable estimation of the covariance matrix is important for calculation of multivariate distance, Ledoit-Wolf shrinkage covariance estimation was employed [51]. This approach provides a regularized covariance estimate and reduces instability associated with inversion of empirical covariance matrices.

For patient $p$, the pairwise deviation vector for morphological characteristics $i$ and $j$ was defined as

$$\delta_{ij}(p) = \left[z_i(p), z_j(p)\right] - \mu_{ij}(B)$$

The disruption weight associated with this feature pair was then calculated using the Mahalanobis distance [52]:

$$d_{ij}^{(p)} = \sqrt{(z_{ij}^{(p)} - \mu_{ij}^{(p)})^T - P_{ij}^{(B)}(z_{ij}^{(p)} - \mu_{ij}^{(B)})}$$

where $P_{ij}^{(B)}$ represents the inverse regularized covariance matrix, or precision matrix, estimated from the benign training distribution of feature pair $i, j$.

The Mahalanobis formulation incorporates both the scale and covariance structure of the two morphological characteristics. Consequently, a large edge weight indicates that the patient's combination of characteristics $i$ and $j$ is unusual relative to the corresponding relationship observed among benign training samples, whereas a small edge weight indicates consistency with the benign feature-pair distribution. The resulting patient-specific morphological disruption network was represented by the symmetric weighted adjacency matrix

$$D(p) = [d_{ij}(p)],\ d_{ii}(p) = 0,\ d_{ij}(p) = d_{ji}(p)$$

Because $K = 10$ morphological characteristics were included, each patient-specific network contained

$$E = \frac{K(K-1)}{2} = 45$$

unique undirected feature-pair relationships. Importantly, the benign pairwise distributions were reconstructed independently within every cross-validation training partition. Therefore, the morphological disruption network of a held-out patient was always calculated relative to benign relationships learned without using that patient's data.

## Development of the MNDI

MNDI was developed to summarize the overall abnormality of the patient-specific morphological disruption network into a single quantitative measure. For patient $p$, MNDI was calculated as the arithmetic mean of the Mahalanobis-based disruption weights [53,54] across all 45 unique morphological feature pairs:

$$MNDI^{(p)} = \frac{1}{E}\sum_{i<j} d_{ij}^{(p)}$$

where $E = 45$ and $d_{ij}(p)$ represents the Mahalanobis-based disruption weight associated with morphological characteristics $i$ and $j$ for patient $p$. MNDI therefore measures the average extent to which a patient's pairwise morphological relationships deviate from the multivariate relationships observed among benign training samples. A lower MNDI indicates that the patient's overall morphological organization is more consistent with the benign reference distributions, whereas a higher MNDI indicates greater disruption across morphological relationships. Unlike a conventional biomarker based on a single nuclear measurement, MNDI integrates information distributed across all 45 pairwise relationships among the ten morphological characteristics. The index therefore provides a patient-level summary of morphological organization rather than representing the magnitude of any single original feature. In addition to MNDI, complementary descriptors of the patient-specific disruption network were calculated from the 45 unique disruption weights. These descriptors included the standard deviation, maximum, minimum, and median disruption weights. The standard deviation characterized heterogeneity in relationship disruption across the network, the maximum represented the most abnormal pairwise relationship, the minimum represented the least disrupted relationship, and the median characterized the central tendency of pairwise disruption. The arithmetic mean of the 45 disruption weights was not included as an additional predictor because it is mathematically identical to MNDI. Accordingly, the final network-derived representation consisted of five nonredundant variables: MNDI, standard deviation of disruption, maximum disruption, minimum disruption, and median disruption.

## Statistical Analysis and Diagnostic Evaluation

The association between MNDI and tumor malignancy was first evaluated by comparing patient-level MNDI values between benign and malignant groups. Because normality of the MNDI distributions was not assumed, differences between diagnostic groups were evaluated using a two-sided Mann-Whitney U test [55]. Statistical significance was defined as

$$\alpha = 0.05$$

The discriminatory performance of MNDI as a standalone biomarker was evaluated using receiver operating characteristic (ROC) analysis. The area under the ROC curve (AUC) was used to quantify the ability of MNDI to distinguish malignant from benign tumors across possible decision thresholds. An optimal operating threshold for MNDI was identified using Youden's index:

$$\mathbf{J = Sensitivity + Specificity - 1}$$

The threshold that maximized J was selected as the optimal operating point, and the corresponding sensitivity and specificity were determined. Ninety-five percent confidence intervals were calculated for principal diagnostic performance estimates in the final analysis. To provide a robust estimate of predictive performance and minimize dependence on a single train-test partition, repeated stratified five-fold cross-validation was used. The dataset was divided into five folds while preserving the relative proportions of benign and malignant cases. Four folds were used for model development and the remaining fold served as held-out validation data. The procedure was repeated ten times using different randomized partitions, resulting in 50 held-out validation folds. All preprocessing and network-construction procedures were performed independently within each cross-validation iteration. Feature-standardization parameters were estimated exclusively from the training folds. The benign feature-pair distributions, including their centers and regularized covariance structures, were then estimated using only benign samples within those training folds. These training-derived parameters were subsequently applied to the held-out fold to construct patient-specific disruption networks and calculate MNDI values. Thus, validation samples did not contribute to feature normalization, benign-reference estimation, network construction parameters, or classifier training. Because each patient appeared once in the held-out validation set during each of the ten cross-validation repetitions, each patient received ten independently derived out-of-fold MNDI estimates. For patient-level statistical and standalone biomarker analyses, these held-out estimates were averaged to obtain a single leakage-free MNDI estimate for each patient. These patient-level estimates were used for the Mann-Whitney U test and standalone ROC analysis. A Random Forest classifier was used to evaluate the predictive information contained in the network-derived representation. The final network model incorporated five nonredundant variables: MNDI, standard deviation of disruption, maximum disruption, minimum disruption, and median disruption. Random Forest models were constructed using 500 decision trees, with a fixed random seed to support reproducibility. Classification performance was evaluated using ROC-AUC, accuracy, sensitivity, specificity, precision, and F1-score. Performance measures were calculated on held-out validation data and summarized across the 50 validation folds as the mean and standard deviation.

To determine how the network representation compared with conventional morphological information, three predictive configurations were evaluated using identical cross-validation partitions. The first model used the ten original mean morphological characteristics and served as the conventional feature-based baseline. The second model used only the five nonredundant network-derived variables. The third model combined the ten original morphological characteristics with the five network-derived variables. In addition, MNDI was evaluated independently as a single-variable biomarker. The use of identical validation partitions enabled direct comparison of conventional morphology, network-derived disruption, and their combination. Permutation-based feature importance was used to examine the predictive contribution of individual network descriptors. Importance was calculated on held-out validation data by independently permuting each network-derived variable and quantifying the resulting reduction in ROC-AUC. This procedure was preferred over impurity-based Random Forest importance because it evaluates the contribution of predictors directly on unseen data and is less dependent on properties of the tree-splitting procedure. Permutation importance estimates were summarized across repeated validation folds. Overall, the proposed framework transforms conventional nuclear morphological measurements into patient-specific disruption networks by quantifying the abnormality of each feature pair relative to benign multivariate relationships. MNDI then summarizes disruption across all 45 pairwise relationships into a single patient-level index. The complete analytical procedure was embedded within repeated cross-validation so that feature normalization, benign-reference estimation, network construction, and predictive evaluation remained separated from held-out validation data.

# Results

## Study Cohort and Network Construction

The study included 569 samples from the Wisconsin Diagnostic Breast Cancer dataset, comprising 357 benign (62.7%) and 212 malignant (37.3%) breast masses. Ten mean nuclear morphological characteristics; radius, texture, perimeter, area, smoothness, compactness, concavity, concave points, symmetry, and fractal dimension; were included in the network analysis. These characteristics constituted the nodes of each patient-specific morphological network. With 10 nodes, each network contained 45 unique undirected feature-pair relationships. Patient-specific disruption weights were calculated for these relationships relative to benign feature-pair distributions derived exclusively from the corresponding training data within the repeated cross-validation framework.

Table 1: Summary of the study cohort and morphological network characteristics.

| Characteristic | Value |
|---|---|
| **Total samples** | 569 |
| **Benign** | 357 (62.7%) |
| **Malignant** | 212 (37.3%) |
| **Morphological characteristics** | 10 |
| **Network nodes** | 10 |
| **Unique feature pairs** | 45 |

## Morphological Network Disruption in Malignant Tumors

Malignant breast masses exhibited markedly higher MNDI values than benign masses. The median MNDI was 3.175 (IQR: 2.317–4.285) in malignant cases compared with 1.066 (IQR: 0.807–1.383) in benign cases. The difference between the two groups was statistically significant (Mann-Whitney U = 4,490, $p = 2.94 \times 10^{-69}$), with a large rank-biserial effect size ($r = 0.881$). Figure 1 presents the distribution of patient-level MNDI values across the two diagnostic groups, illustrating the pronounced shift toward higher network disruption values among malignant cases.

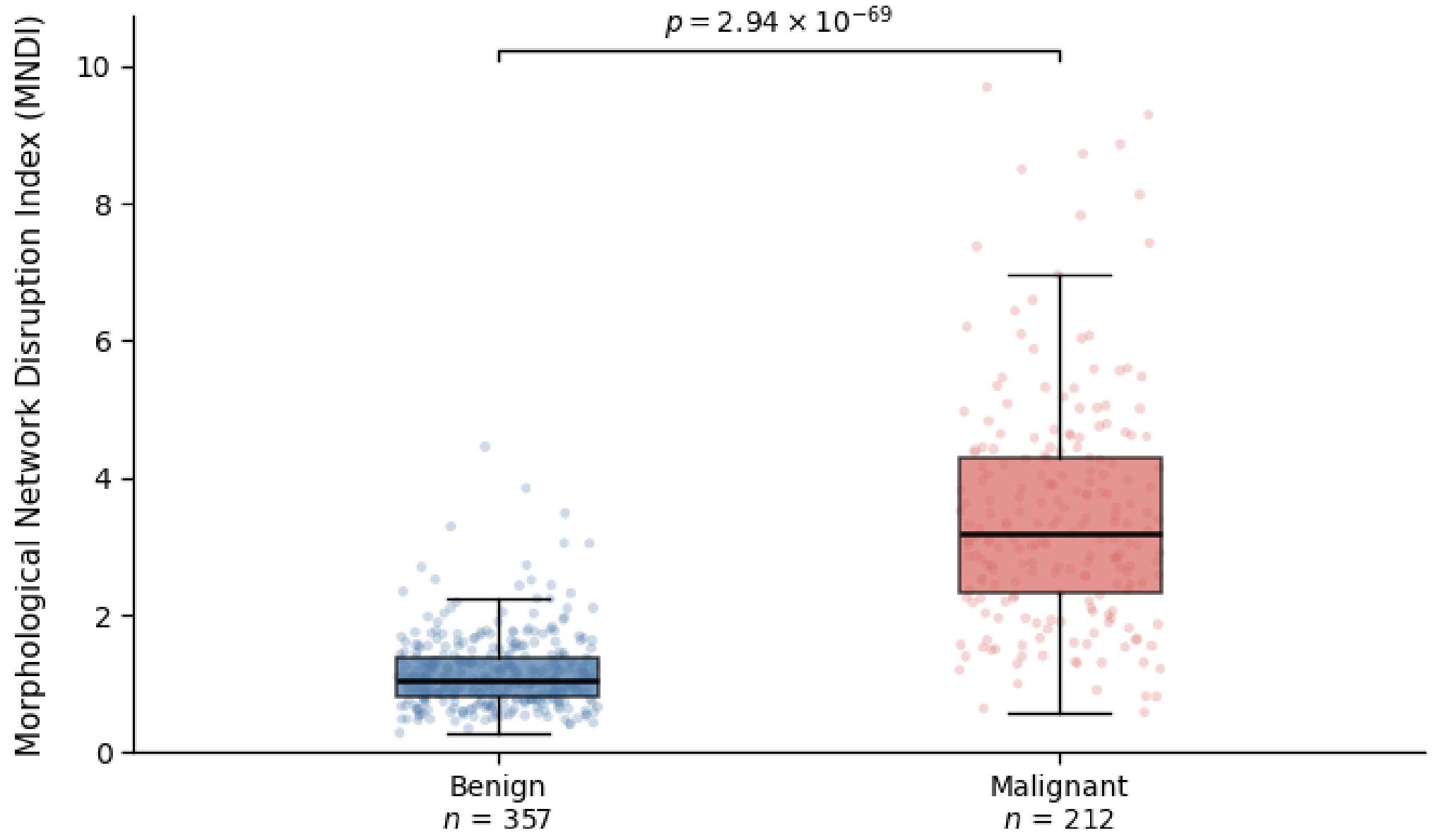

Figure 1: Distribution of the Morphological Network Disruption Index in benign and malignant breast masses.

## Diagnostic Performance of MNDI as a Standalone Biomarker

The MNDI demonstrated strong discriminatory performance for distinguishing malignant from benign breast masses, achieving a ROC-AUC of 0.941 (95% bootstrap CI: 0.918–0.961). The optimal threshold determined by the Youden index was 1.797, corresponding to a sensitivity of 85.9%, a specificity of 91.6%, and a Youden index of 0.775. Figure 2a presents the receiver operating characteristic curve and the corresponding optimal operating point, whereas Figure 2b shows the variation in sensitivity and specificity across MNDI thresholds and identifies the optimal threshold of 1.797.

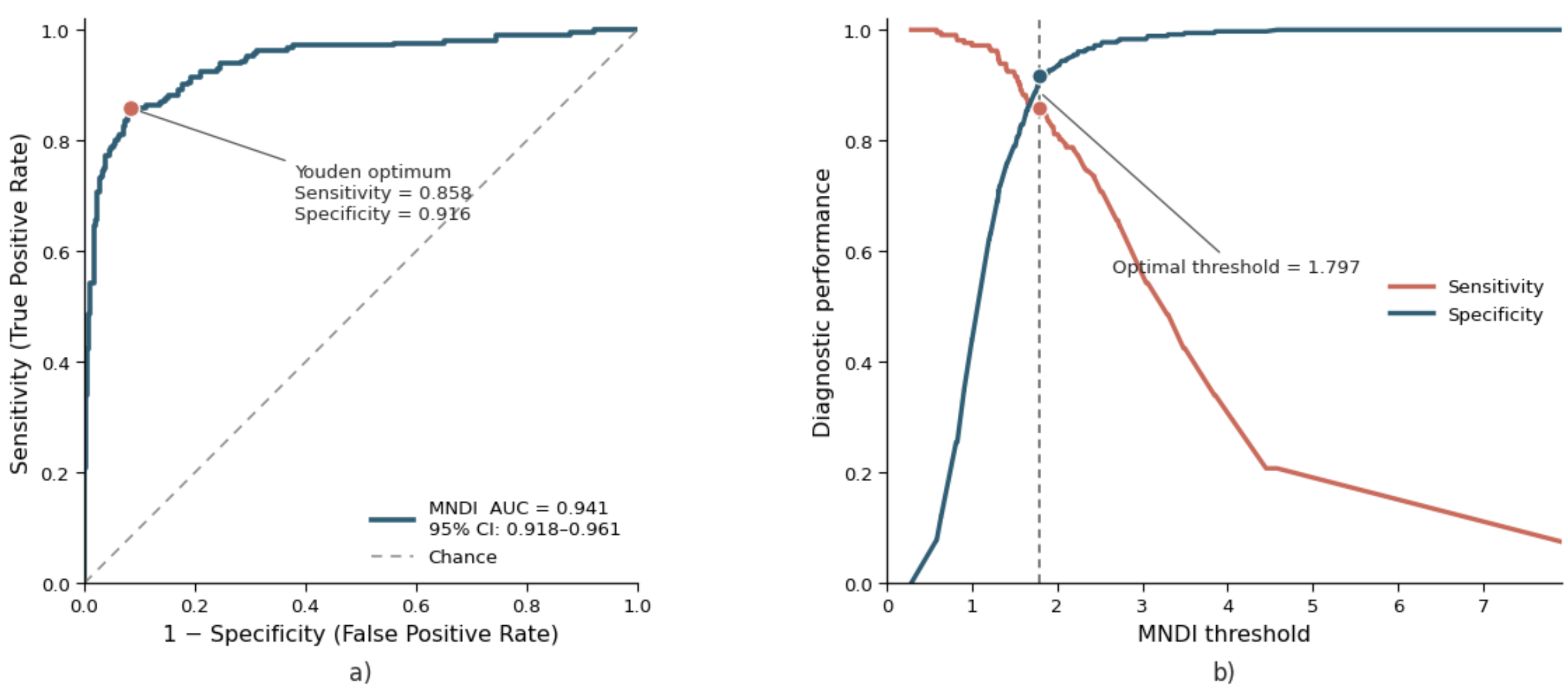

Figure 2: Diagnostic performance of MNDI. (a) ROC curve with the Youden-optimal operating point. (b) Sensitivity and specificity across MNDI thresholds.

### Predictive Performance of Network-Derived Features

The network-derived feature model demonstrated consistent performance across repeated cross-validation, achieving a mean ROC-AUC of 0.928, accuracy of 0.869, sensitivity of 0.793, specificity of 0.914, precision of 0.849, and F1-score of 0.817. Figure 3a shows the distributions of these performance metrics across the 50 held-out validation folds. Held-out permutation analysis showed that median disruption had the highest importance, followed by MNDI and maximum disruption, whereas SD and minimum disruption showed comparatively smaller contributions to model performance (Figure 3b).

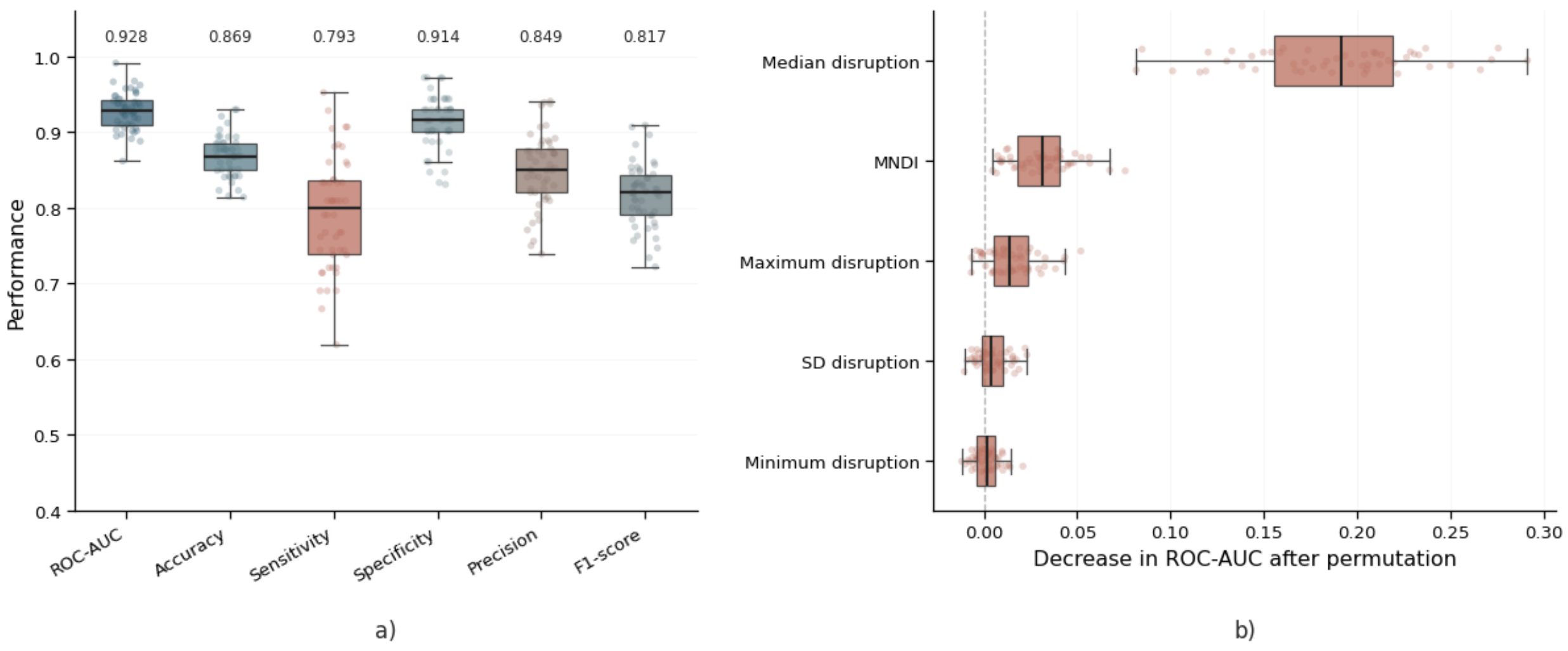

Figure 3: ross-validated performance and permutation importance of the network-derived model. (a) Distribution of classification metrics across 50 held-out validation folds. (b) Held-out permutation importance of the five network descriptors, expressed as the decrease in ROC-AUC following permutation.

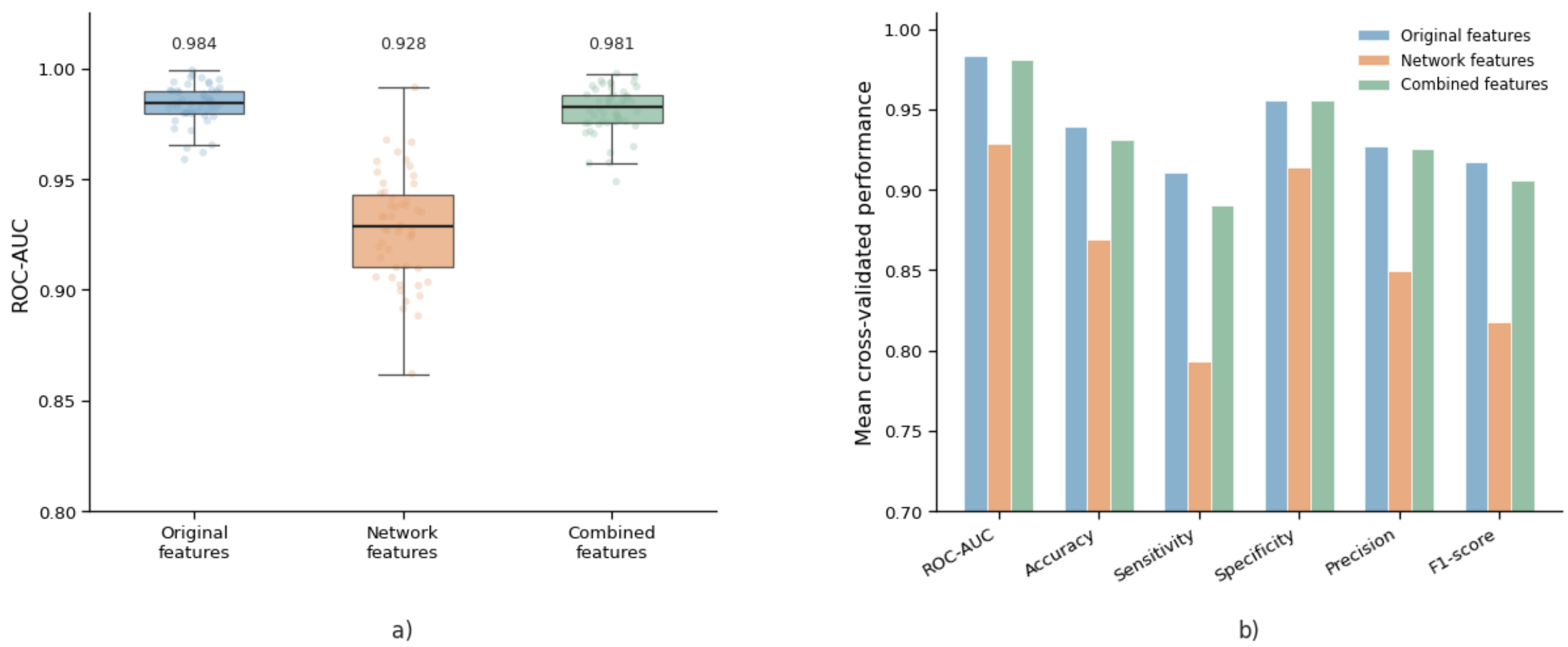

Figure 4: Comparison of predictive performance across feature representations. (a) ROC-AUC distributions across 50 held-out validation folds for the original, network-derived, and combined feature models. (b) Mean cross-validated performance across six classification metrics for the three models.

### Comparison With Conventional Morphological Features

The predictive performance of the network-derived features was compared with that of the original morphological features and their combination (Table 2). The original feature model achieved a mean ROC-AUC of 0.984 ± 0.009, compared with 0.928 ± 0.024 for the network-derived model and 0.981 ± 0.011 for the combined model. The corresponding accuracies were 0.939 ± 0.025, 0.869 ± 0.027, and 0.931 ± 0.025, respectively. The original feature model also achieved a sensitivity of 0.910 ± 0.053, specificity of 0.956 ± 0.030, precision of 0.927 ± 0.046, and F1-score of 0.917 ± 0.035. The combined model produced comparable performance, with a sensitivity of 0.890 ± 0.061, specificity of 0.956 ± 0.029, precision of 0.925 ± 0.045, and F1-score of 0.906 ± 0.036. Figure 4 shows the distributions of ROC-AUC across the held-out folds and the overall performance profiles of the three models.

Table 2: Cross-validated predictive performance of the three feature models

| Metric | Original features | Network features | Combined features |
|---|---|---|---|
| **ROC-AUC** | **0.984 ± 0.009** | 0.928 ± 0.024 | 0.981 ± 0.011 |
| **Accuracy** | **0.939 ± 0.025** | 0.869 ± 0.027 | 0.931 ± 0.025 |
| **Sensitivity** | **0.910 ± 0.053** | 0.793 ± 0.073 | 0.890 ± 0.061 |
| **Specificity** | **0.956 ± 0.030** | 0.914 ± 0.034 | 0.956 ± 0.029 |
| **Precision** | **0.927 ± 0.046** | 0.849 ± 0.048 | 0.925 ± 0.045 |
| **F1-score** | **0.917 ± 0.035** | 0.817 ± 0.041 | 0.906 ± 0.036 |

### Pairwise Morphological Disruption Associated with Malignancy

Pairwise analysis of the morphological network showed greater disruption in malignant than benign breast masses across all 45 feature pairs. All 45 pairwise comparisons remained statistically significant after Benjamini-Hochberg false discovery rate (FDR) correction [56]. The largest effect was observed for the area–concave points pair, with median disruption values of 4.552 in malignant and 1.129 in benign cases and a rank-biserial effect size of 0.919 (FDR-adjusted $p = 1.93 \times 10^{-73}$). This was followed by radius–concave points ($r = 0.909$), perimeter-concave points ($r = 0.908$), and area–concavity ($r = 0.901$). Several additional pairwise configurations involving concavity, concave points, compactness, and measures of nuclear size also showed large effect sizes (Table 3). Figure 5a summarizes the effect sizes across the complete set of 45 morphological feature pairs, while Figure 5b presents the 10 pairs with the largest rank-biserial effect sizes.

Table 3: Morphological feature pairs with the largest malignancy-associated disruption

| Morphological feature pair | Benign median | Malignant median | Rank-biserial $r$ | FDR-adjusted $p$ |
|---|---|---|---|---|
| **Area – Concave points** | 1.129 | 4.552 | **0.919** | $1.93 \times 10^{-73}$ |
| **Radius – Concave points** | 1.134 | 4.175 | **0.909** | $2.80 \times 10^{-72}$ |
| **Perimeter – Concave points** | 1.115 | 4.181 | **0.908** | $2.82 \times 10^{-72}$ |
| **Area – Concavity** | 0.968 | 4.327 | **0.901** | $3.01 \times 10^{-71}$ |
| **Perimeter – Concavity** | 0.963 | 4.145 | **0.893** | $4.41 \times 10^{-70}$ |
| **Radius – Concavity** | 0.977 | 4.046 | **0.892** | $5.01 \times 10^{-70}$ |
| **Area – Compactness** | 1.122 | 4.135 | **0.885** | $5.20 \times 10^{-69}$ |
| **Compactness – Concave points** | 1.055 | 4.080 | **0.885** | $5.20 \times 10^{-69}$ |
| **Texture – Concave points** | 1.012 | 3.988 | **0.883** | $7.79 \times 10^{-69}$ |
| **Concave points – Fractal dimension** | 0.982 | 3.798 | **0.879** | $2.68 \times 10^{-68}$ |

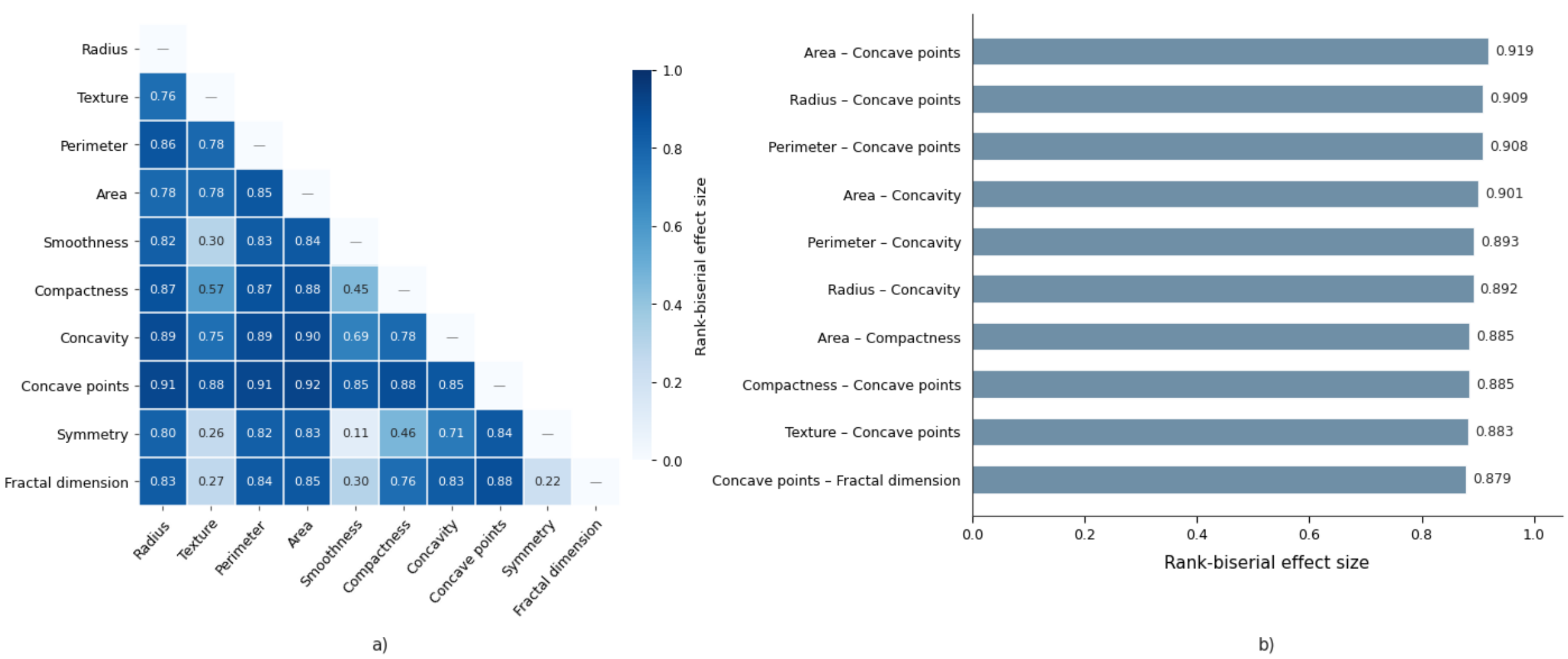

Figure 5: Pairwise morphological disruption associated with malignancy. (a) Effect sizes across 45 morphological feature pairs. (b) Top 10 pairs with the largest effect sizes.

## Stability of MNDI Across Cross-Validation Repetitions

Each patient received 10 held-out MNDI estimates across the repeated cross-validation procedure. MNDI showed low relative within-patient variability, with a median SD of 0.0268 (IQR: 0.0187–0.0456) and a median coefficient of variation (CV) of 0.0190 (IQR: 0.0153–0.0231). The median range across repeated estimates was 0.0860. Although the median SD was higher in malignant than benign cases (0.0529 vs. 0.0207), relative variability remained low in both groups, with median CVs of 0.0175 and 0.0200, respectively (Table 4). Figure 6 shows the relationship between mean patient-level MNDI and within-patient variability.

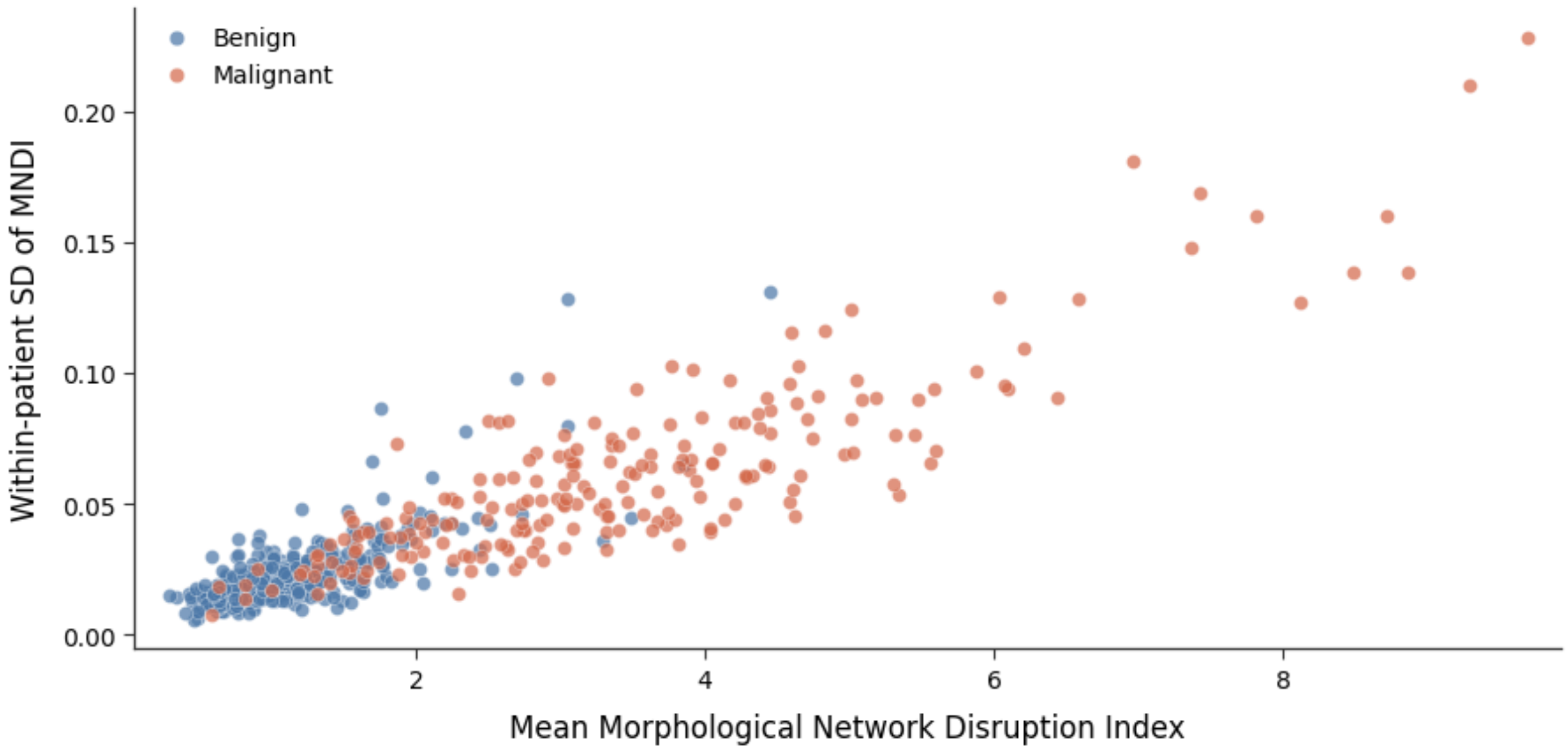

Figure 6: Stability of MNDI across repeated cross-validation. Mean MNDI versus within-patient SD across 10 held-out estimates.

Table 4: Stability of MNDI across repeated cross-validation.

| Stability measure | Overall (n = 569) | Benign (n = 357) | Malignant (n = 212) |
|---|---|---|---|
| **Within-patient SD, median (IQR)** | 0.0268 (0.0187–0.0456) | 0.0207 (0.0157–0.0271) | 0.0529 (0.0388–0.0756) |
| **CV, median (IQR)** | 0.0190 (0.0153–0.0231) | 0.0200 (0.0163–0.0243) | 0.0175 (0.0141–0.0212) |

## External Validation

To evaluate the generalizability of the MNDI framework in an independent breast tumor cohort, the BreaKHis histopathology dataset was analyzed. The dataset included 7,909 H&E-stained breast biopsy images obtained from 82 patients, comprising 24 patients with benign lesions and 58 patients with malignant tumors. Images were acquired at four magnification levels (40×, 100×, 200×, and 400×), with 2,480 benign and 5,429 malignant images available for analysis. Representative benign and malignant histopathology images across the four magnification levels are shown in Figure 7. Patient identifiers were retained throughout the analysis to ensure that images originating from the same patient remained grouped during subsequent validation procedures.

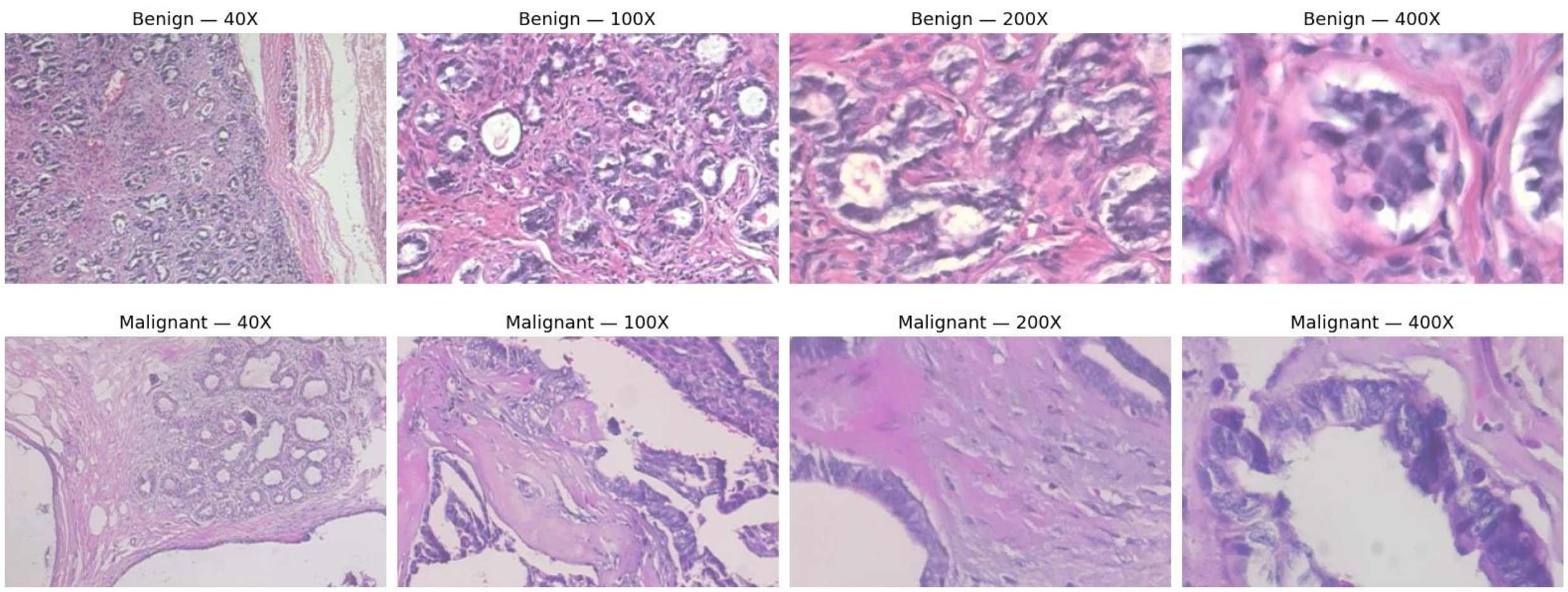

Figure 7: Representative BreaKHis histopathology images. Representative benign and malignant breast biopsy images acquired at 40×, 100×, 200×, and 400× magnification.

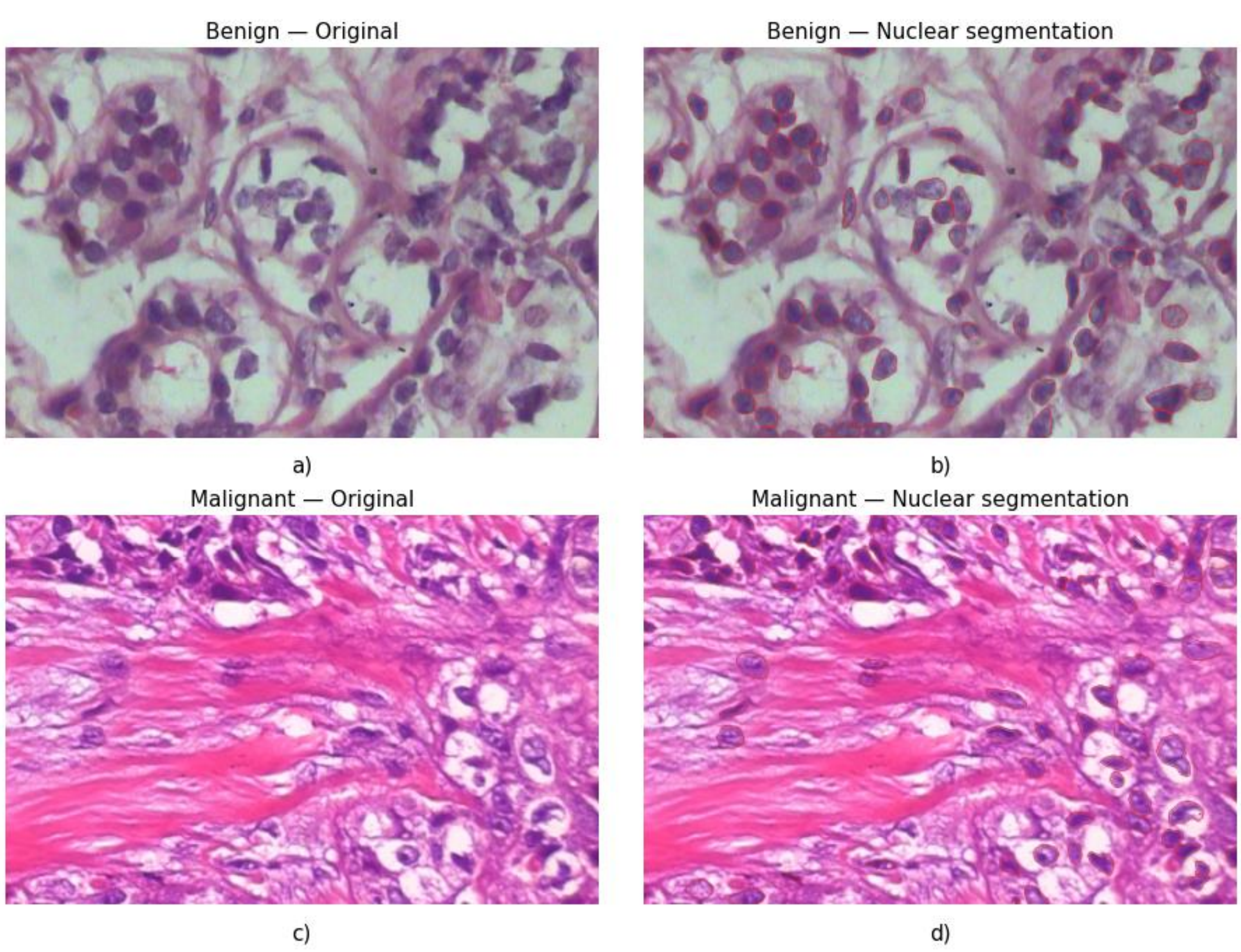

Figure 8: Nuclear segmentation in the BreaKHis cohort. Representative 400× H&E images of benign and malignant breast lesions and corresponding nuclear segmentation overlays.

Nuclear segmentation was performed on the H&E-stained images to enable quantitative characterization of nuclear morphology. A pretrained Cellpose segmentation model [57,58] was applied to identify individual nuclei, and the resulting instance masks were used to delineate nuclear boundaries. Visual quality assessment across images from benign and malignant lesions showed that the segmentation generally captured individual nuclei when nuclear structures were clearly visible, although the number of detected nuclei varied according to tissue composition and image content. Representative nuclear segmentation results are shown in Figure 8. Following nuclear segmentation, image-level quality control was applied to exclude images with insufficient nuclear content. Segmented objects touching the image boundaries and objects outside the predefined size range were excluded [59]. Images containing at least 20 retained nuclei were considered suitable for subsequent morphological analysis. This criterion retained 1,462 of 1,820 (80.3%) 400× images, including 527 benign and 935 malignant images, while preserving all 82 patients in the independent cohort (24 benign and 58 malignant).

Quantitative nuclear morphology was extracted from the retained segmented nuclei and aggregated first at the image level and subsequently at the patient level. To construct the BreaKHis morphological network while minimizing mathematical redundancy among the extracted descriptors, five nuclear morphology characteristics representing complementary aspects of nuclear size and shape were retained: area, perimeter, circularity, concavity, and eccentricity. These five characteristics formed the network nodes, yielding 10 unique pairwise feature combinations for each patient. The BreaKHis network therefore represented an independent implementation of the MNDI framework using histopathology-derived nuclear morphology rather than a direct replication of the WDBC feature set.

Patient-level MNDI values were calculated using repeated stratified five-fold cross-validation with 10 repetitions, with feature standardization and benign reference distributions estimated exclusively from the corresponding training folds. Each patient received 10 held-out MNDI estimates, which were averaged to obtain a single patient-level value. Malignant lesions showed a slightly higher MNDI than benign lesions (mean ± SD: 1.466 ± 0.733 vs. 1.299 ± 0.582; median [IQR]: 1.392 [0.904–1.863] vs. 1.217 [1.027–1.642]); however, the difference was not statistically significant (Mann-Whitney U = 749, p = 0.593; rank-biserial r = 0.076). The MNDI yielded a ROC-AUC of 0.538 (95% bootstrap CI: 0.401–0.668). Figure 9 presents the patient-level MNDI distributions and corresponding receiver operating characteristic curve. At the exploratory Youden threshold of 1.745, sensitivity was 34.5% and specificity was 83.3%. A sensitivity analysis restricted to patients contributing at least five qualifying images produced similar results (n = 75; ROC-AUC = 0.545, 95% bootstrap CI: 0.406–0.683; p = 0.539; rank-biserial r = 0.090). The results of the primary and sensitivity analyses are summarized in Table 5.

Table 5: Independent evaluation of the MNDI framework in the BreaKHis cohort

| Analysis | Patients (B/M) | Benign MNDI, mean | Malignant MNDI, mean | ROC-AUC (95% CI) | p-value | Rank-biserial r |
|---|---|---|---|---|---|---|
| **Primary analysis** | 82 (24/58) | 1.299 | 1.466 | 0.538 (0.401–0.668) | 0.593 | 0.076 |
| **≥5-image sensitivity analysis** | 75 (23/52) | 1.292 | 1.464 | 0.545 (0.406–0.683) | 0.539 | 0.090 |

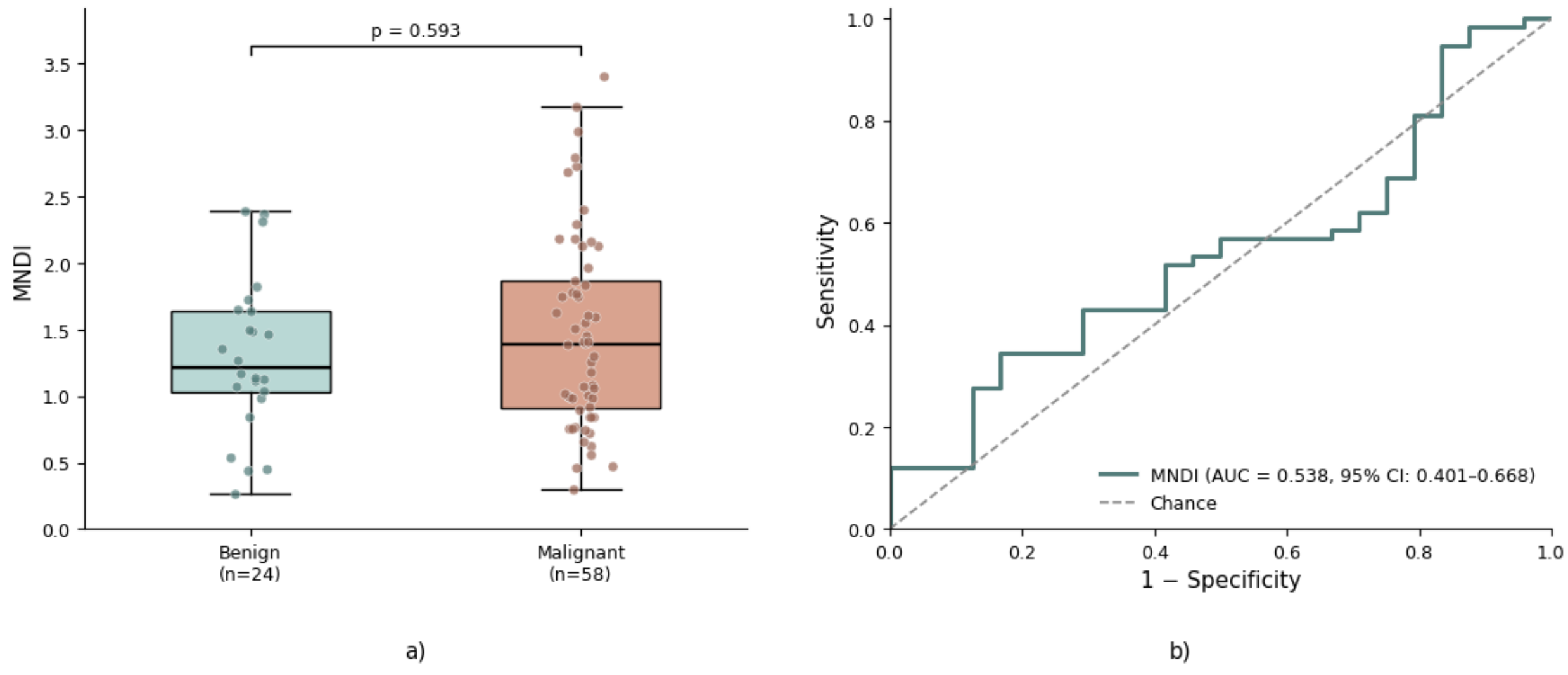

Figure 9: Patient-level MNDI performance in the BreaKHis cohort. (a) MNDI distributions in benign and malignant lesions. (b) ROC curve for MNDI-based discrimination

# Discussion

This study introduced the Morphological Network Disruption Index (MNDI), a patient-level measure of morphological abnormality defined relative to a benign reference state. Rather than treating nuclear characteristics exclusively as independent predictors, MNDI summarizes the extent to which pairwise morphological configurations in an individual lesion depart from benign joint feature distributions. The principal finding was a marked increase in morphological disruption among malignant breast masses in WDBC. MNDI achieved an ROC-AUC of 0.941 (95% CI: 0.918–0.961), while a classifier based on a small set of network-derived disruption summaries achieved a mean cross-validated ROC-AUC of 0.928. These findings indicate that substantial diagnostic information can be retained when conventional morphological measurements are reformulated as patient-specific deviations from a reference morphological state.

The magnitude of the WDBC difference was notable. Median MNDI increased from 1.066 in benign lesions to 3.175 in malignant lesions, with a large rank-biserial effect size ($r = 0.881$). Importantly, this separation was obtained under repeated held-out validation: feature standardization and benign reference distributions were constructed exclusively from training samples, and each patient was evaluated against a reference distribution to which that patient had not contributed. This design is particularly important for reference-based biomarkers because constructing the reference state using test observations could artificially influence the estimated degree of abnormality.

The pairwise analysis provided additional insight into the morphological basis of the MNDI signal. The strongest malignancy-associated disruptions involved combinations of nuclear size measures; particularly area, radius, and perimeter; with concave points and concavity. Area-concave points showed the largest effect, followed by radius-concave points and perimeter-concave points. Thus, the global increase in MNDI was not distributed uniformly across the morphological feature space but was concentrated in configurations combining nuclear enlargement with boundary irregularity. This decomposition is an important feature of the framework: MNDI provides a single patient-level summary, while the underlying pairwise disruption profile identifies which morphological configurations contribute most strongly to that abnormality.

MNDI should nevertheless be interpreted as a measure of joint morphological abnormality, rather than as evidence of altered biological interactions between individual features. The pairwise Mahalanobis distance reflects displacement within a two-dimensional feature distribution and can therefore increase because of changes in individual feature magnitudes, their joint configuration, or both. Accordingly, the network edges used here represent statistical morphological configurations rather than mechanistic interactions. This distinction provides a more precise interpretation of the proposed network and avoids attributing biological causality to statistical relationships.

The comparison with conventional features further clarifies the contribution of the approach. Original WDBC morphological features achieved a mean ROC-AUC of 0.984, exceeding the network-derived model (0.928), while combining both representations did not improve performance (ROC-AUC = 0.981). MNDI should therefore not be viewed as a replacement for conventional features when maximizing classification accuracy is the sole objective. Its contribution is instead representational: a multidimensional morphological profile is transformed into an interpretable patient-level measure describing the magnitude of deviation from a benign reference state. The relatively low within-patient variability of MNDI across repeated held-out estimates further suggests that this representation was not highly dependent on a particular cross-validation partition, although this should not be interpreted as evidence of longitudinal or test-retest reproducibility.

The BreaKHis analysis provided an important test of what is and is not generalizable about this framework. Because BreaKHis contains H&E-stained tissue sections rather than the FNA-derived measurements available in WDBC, the original WDBC model could not simply be transferred unchanged. Instead, the underlying MNDI principle was reconstructed using independently segmented nuclei and five nonredundant histopathology-derived descriptors representing nuclear size and shape. This distinction is important: the BreaKHis experiment evaluated generalization of the framework across morphological representations, rather than external validation of an identical frozen WDBC predictor. The resulting BreaKHis MNDI showed limited discrimination (ROC-AUC = 0.538, 95% CI: 0.401–0.668), and the difference between benign and malignant lesions was not significant. Restricting the analysis to patients represented by at least five qualifying images produced a nearly identical result (ROC-AUC = 0.545), indicating that the finding was not explained by patients with sparse image representation. Although this result did not reproduce the diagnostic separation observed in WDBC, it provides an informative boundary condition for MNDI. Specifically, it demonstrates that the mathematical operation of measuring deviation from a benign reference is not, by itself, sufficient to generate a discriminative biomarker when the underlying morphological representation changes substantially.

This observation strengthens an important conceptual aspect of the proposed approach. MNDI is best understood as a framework for quantifying disruption within a defined morphological state space, rather than as a universal numerical score that should remain invariant across modalities. The meaning of a Mahalanobis-based deviation depends on the variables defining that state space and on the reference population used to estimate its geometry. WDBC and BreaKHis differ not only in patient populations but also in specimen preparation, image acquisition, cellular context, feature definitions, and measurement procedures. Consequently, a five-node network reconstructed from segmented H&E nuclei does not represent the same morphological state space as the WDBC network. The contrast between the two datasets therefore helps distinguish generalizability of the MNDI principle from transportability of a specific MNDI biomarker.

This distinction also identifies a concrete direction for future development. The BreaKHis implementation was intentionally based on a limited set of geometric nuclear descriptors, including area, perimeter, circularity, concavity, and eccentricity. Histopathology contains substantially richer information, including chromatin texture, staining heterogeneity, pleomorphism, nuclear spatial organization, cell density, and tissue architecture. The absence of strong discrimination in BreaKHis suggests that an effective histopathology-specific MNDI may require a feature space capable of representing these additional dimensions rather than simply transferring a geometric network developed in another morphological context. Importantly, such extensions should be defined prospectively and evaluated in independent data rather than selected retrospectively to maximize performance in the present cohort.

## Limitation

Several limitations should be acknowledged. WDBC remains the primary dataset, and the strong MNDI performance requires confirmation in an independent cohort with compatible morphological measurements. The WDBC threshold of 1.797 was data-derived and should therefore be considered exploratory rather than a clinically validated cutoff. In addition, the pairwise formulation does not explicitly capture higher-order relationships among three or more morphological characteristics, and the resulting distances quantify joint abnormality rather than isolated changes in statistical dependence. In the BreaKHis analysis, the relatively small number of patients, particularly the 24 benign cases, limited the precision of the benign reference distributions. Automated nuclear segmentation and the restricted set of geometric descriptors also introduced sources of measurement variability that were not present in the original WDBC representation.

Future studies should address two distinct validation objectives. First, the WDBC-derived MNDI should be evaluated in an independent cohort with harmonized nuclear morphological measurements, allowing the feature definitions, reference construction procedure, and diagnostic operating point to be specified before validation. Second, histopathology-specific implementations should incorporate richer representations of nuclear and tissue morphology, including texture and spatial organization, and subsequently be evaluated in independent cohorts. Distinguishing validation of a fixed MNDI biomarker from extension of the underlying MNDI framework will be important for establishing both biomarker transportability and methodological generalizability.

## Conclusion

MNDI provides a compact, patient-specific representation of morphological deviation from a benign reference state. In WDBC, MNDI demonstrated strong discrimination between benign and malignant lesions, low variability across repeated held-out estimates, and identifiable morphological configurations contributing to malignancy-associated disruption. The BreaKHis analysis further clarified an important property of the approach: MNDI is not a feature-agnostic score, but a framework whose meaning depends on the morphological state space in which disruption is measured. Thus, the transferable concept is the reference-based quantification of patient-specific morphological disruption, whereas individual MNDI implementations require validation within biologically and technically compatible feature spaces. This distinction provides a foundation for developing interpretable, modality-specific network biomarkers of morphological abnormality.

**Data Availability Statement**

The datasets analyzed in this study are publicly available. The Wisconsin Diagnostic Breast Cancer (WDBC) dataset is available through the UCI Machine Learning Repository. The BreaKHis histopathology dataset is publicly available through its original repository and associated public data distributions. No new patient data were collected in this study.

**Code Availability**

The code used for data preprocessing, morphological network construction, calculation of the Morphological Network Disruption Index (MNDI), statistical analysis, and figure generation will be made available upon reasonable request.

**Ethics Approval and Consent to Participate**

This study involved secondary analysis of publicly available, de-identified datasets and did not involve direct interaction with human participants or the collection of new patient data. Therefore, institutional review board approval and additional informed consent were not required for the present study.

**Consent for Publication**

Not applicable.

**Competing Interests**

The author declares no competing interests.

**Funding**

This research received no specific grant from any funding agency in the public, commercial, or not-for-profit sectors.

**Author Contributions**

The author was solely responsible for the conceptualization, methodology, software implementation, data analysis, validation, visualization, interpretation of results, and preparation and revision of the manuscript.

**Acknowledgments**

The author acknowledges the investigators and institutions responsible for developing and publicly providing the Wisconsin Diagnostic Breast Cancer (WDBC) and BreaKHis datasets used in this study.